\documentclass[12pt]{revtex4}
\usepackage{amssymb}

\begin{document}

\title{Horizonless Rotating Solutions in $(n+1)$-dimensional Einstein-Maxwell
Gravity}
\author{M. H. Dehghani}\email{dehghani@physics.susc.ac.ir}
\address{Physics Department and Biruni Observatory,
         Shiraz University, Shiraz 71454, Iran\\ and\\
         Institute for Studies in Theoretical Physics and Mathematics (IPM)\\
         P.O. Box 19395-5531, Tehran, Iran}

\begin{abstract}
We introduce two classes of rotating solutions of Einstein-Maxwell
gravity in $n+1$ dimensions which are asymptotically anti-de
Sitter type. They have no curvature singularity and no horizons.
The first class of solutions, which has a conic singularity yields
a spacetime with a longitudinal magnetic field and $k$ rotation
parameters. We show that when one or more of the rotation
parameters are non zero, the spinning brane has a net electric
charge that is proportional to the magnitude of the rotation
parameters. The second
class of solutions yields a spacetime with an angular magnetic field and $%
\kappa$ boost parameters. We find that the net electric charge of
these traveling branes with one or more nonzero boost parameters
is proportional to the magnitude of the velocity of the brane. We
also use the counterterm method inspired by AdS/CFT correspondence
and calculate the conserved quantities of the solutions. We show
that the logarithmic divergencies associated to the Weyl anomalies
and matter field are zero, and the $r$ divergence of the action
can be removed by the counterterm method.
\end{abstract}

\maketitle


\section{Introduction}

The anti-de Sitter conformal field theory (AdS/CFT) correspondence
\cite{Mal}, which relates the low energy limit of string theory in
asymptotically anti de-Sitter (AAdS) spacetime and the quantum
field theory living on the boundary of it, have attracted a great
deal of attention in recent years. This equivalence between the
two formulations means that, at least in principle, one can obtain
complete information on one side of the duality by performing
computation on the other side. A dictionary translating between
different quantities in the bulk gravity theory and its
counterparts on the boundary has emerged, including the partition
functions of both theories. An interesting application of the
AdS/CFT correspondence is the interpretation of the Hawking-Page
phase transition between a thermal AdS and AAdS black hole as the
confinement-deconfinement phases of the Yang-Mills (dual gauge)
theory defined on the AdS boundary \cite{Wit}. This conjecture is
now a fundamental concept that furnishes a means for calculating
the action and conserved quantities intrinsically without reliance
on any reference spacetime \cite {Sken1,BK,Od1}. It has also been
applied to the case of black holes with constant negative or zero
curvature horizons \cite{Deh1} and rotating higher genus black
branes \cite{Deh2}. Although the AdS/CFT correspondence applies
for the case of a spatially infinite boundary, it was also
employed for the computation of the conserved and thermodynamic
quantities in the case of a finite boundary \cite{Deh3}. The
counterterm method also has been extended to the case of
asymptotically de Sitter spacetimes \cite{dS}.

Thus, it has become important to obtain new solutions of the
Einstein gravity with a negative cosmological constant and apply
the AdS/CFT correspondence to them. For AAdS spacetimes, the
presence of a negative cosmological constant makes it possible to
have a large variety of static and stationary solutions with
planar, spherical, or cylindrical symmetries. Asymptotically AdS
black holes whose event horizons are hypersurfaces with a
positive, negative, or zero scalar curvature were considered in
\cite{Man1}. Asymptotically AdS rotating topological black brane
solutions have also been obtained \cite{Klem}. The AAdS rotating
solution of Einstein's equation with cylindrical and toroidal
horizons and its extension to include the electromagnetic field
were considered in Ref. \cite{Lem1}. The generalization of this
AAdS charged rotating solution of the Einstein-Maxwell equation to
higher dimensions and higher derivative gravity and investigation
of its thermodynamics were done in \cite{Awad,Deh4}. In the
context of gauged supergravity models, the rotating brane solution
has been considered by many authors (see, for example,
\cite{Cvet}).

In this paper we are dealing with the issue of the spacetimes
generated by spinning string/brane sources in $(n+1)$-dimensional
Einstein-Maxwell theory that are horizonless and have nontrivial
external solutions. These kinds of solutions have been
investigated by many authors in four dimensions. Static uncharged
cylindrically symmetric solutions of Einstein gravity in four
dimensions were considered in \cite{Levi}. Similar static
solutions in the context of cosmic string theory were found in
\cite{Vil}. All of these solutions \cite{Levi,Vil} are horizonless
and have a conical geometry; they are everywhere flat except at
the location of the line source. The extension to include the
electromagnetic field has also been done \cite {Muk,Lem2}. Some
solutions of type IIB supergravity compactified on a four
dimensional torus were considered in \cite{Lun}, which have no
curvature singularity and no conic singularity. Here we will
generalize the
four-dimensional solution found in \cite{Lem2} to the case of the $(n+1)$%
-dimensional solution with all rotation and boost parameters, and use the
AdS/CFT correspondence to compute the conserved quantities of the system.

The outline of our paper is as follows. In Sec. \ref{Sol}, we
first review the field equations of Einstein-Maxwell gravity and
then introduce two classes of $(n+1)$-dimensional AAdS horizonless
charged rotating solutions with more rotation and boost
parameters. In Sec. \ref{Cons}, we give a brief review of AdS/CFT
correspondence, and obtain the logarithmic divergences associated
with the Weyl anomalies and matter fields. The electric charge and
conserved quantities of the system will also be computed. We
finish our paper with some concluding remarks.

\section{(n+1)-dimensional rotating solutions of Einstein-Maxwell gravity
\label{Sol}}

\subsection{Field equations}

The gravitational action for Einstein-Maxwell theory in $n+1$
dimensions for AAdS spacetimes is
\begin{equation}
I_{G}=-\frac{1}{16\pi }\int_{\mathcal{M}}d^{n+1}x\sqrt{_{-}g}\left(
R-2\Lambda -F^{\mu \nu }F_{\mu \nu }\right) +\frac{1}{8\pi }\int_{\partial
\mathcal{M}}d^{n}x\sqrt{_{-}\gamma }K(\gamma ),  \label{Actg}
\end{equation}
where $F_{\mu \nu }=\partial _{\mu }A_{\nu }-\partial _{\nu
}A_{\mu }$ is the electromagnetic tensor field and$\ A_{\mu }$ is
the vector potential. The first term is the Einstein-Hilbert
volume term with negative cosmological constant $\Lambda
=-n(n-1)/2l^{2}$ and the second term is the Gibbons-Hawking
boundary term which is chosen such that the variational principle
is well defined. The manifold $\mathcal{M}$ has metric $g_{\mu \nu
}$ and covariant derivative $\nabla _{\mu }$. $K$ is the trace of
the extrinsic curvature $K^{\mu \nu }$ of any boundary(ies)
$\partial \mathcal{M} $ of the manifold $\mathcal{M}$, with
induced metric(s) $\gamma _{i,j}$.

Varying the action over the metric tensor $g_{\mu \nu }$ and electromagnetic
field $F_{\mu \nu }$, the equations of gravitational and electromagnetic
fields are obtained as
\begin{equation}
\ R_{\mu \nu }-\frac{1}{2}g_{\mu \nu }R-\frac{n(n-1)}{2l^{2}}g_{\mu \nu
}=T_{\mu \nu },  \label{Geq}
\end{equation}
\begin{equation}
\nabla _{\mu }F_{\mu \nu }=0,  \label{EMeq}
\end{equation}
where $T_{\mu \nu }^{em}$ is the electromagnetic stress tensor
\begin{equation}
T_{\mu \nu }^{em}=2{F^{\lambda }\ _{\mu }}F_{\lambda \nu }-\frac{1}{2}%
F_{\lambda \sigma }F^{\lambda \sigma }g_{\mu \nu }.  \label{Str}
\end{equation}

Now we want to obtain spacetimes generated by brane sources in
$(n+1)$ dimensions which satisfy the above field equations. Some
solutions were obtained in \cite{Sab}. We will introduce two new
classes of horizonless solutions to Eqs. (\ref{Geq})-(\ref{Str}),
which are a generalization of those given by Dias and Lemos
\cite{Lem2}.

\subsection{Longitudinal magnetic field solutions with one rotation parameter%
\label{Lmag1}}

It is a matter of calculation to show that the generalization of
the four-dimensional solution given by Dias and Lemos \cite{Lem2}
in $n+1$ dimensions with one rotation parameter can be written as

\begin{equation}
ds^{2} =-\frac{\rho ^{2}}{l^{2}}\left( \Xi dt-ad\phi \right)
^{2}+f(\rho )\left( \frac{a}{l}dt-\Xi ld\phi \right) ^{2}
+\frac{d\rho ^{2}}{f(\rho )}+\frac{\rho ^{2}}{l^{2}}dX^{2},
\label{Metr1a}
\end{equation}
where $\Xi =\sqrt{1+a^{2}/l^{2}}$,
$dX^{2}=\sum_{i=1}^{n-2}(dx^{i})^{2}$ is
the Euclidean metric on the $\left(n-2\right)$-dimensional submanifold, and $%
f(\rho )$ is
\begin{equation}
f(\rho )=\frac{\rho ^{2}}{l^{2}}+\frac{8ml^{n-2}}{\rho ^{n-2}}-\frac{8}{%
(n-1)(n-2)}\frac{q^{2}l^{2n-4}}{\rho ^{2n-4}}.  \label{Ff1}
\end{equation}
The gauge potential is given by

\begin{equation}
A_{\mu }=\frac{2}{(n-2)}\frac{ql^{(n-3)}}{\rho ^{n-2}}\left( a\delta _{\mu
}^{0}-\Xi l^{2}\delta _{\mu }^{\phi }\right) .  \label{pot1}
\end{equation}
As we will see in Sec. \ref{Cons}, $m$ and $q$ in Eqs.
(\ref{Metr1a})-(\ref {pot1}) are the mass and charge parameters of
the metric which are related to the mass and charge densities of
the brane.

In order to study the general structure of this solution, we first look for
curvature singularities. It is easy to show that the Kretschmann scalar $%
R_{\mu \nu \lambda \kappa }R^{\mu \nu \lambda \kappa }$ diverges
at $\rho =0$ and therefore one might think that there is a
curvature singularity located at $\rho =0$. However, as will be
seen below, the spacetime will never achieve $\rho =0$. Now we
look for the existence of horizons, and therefore we look for
possible black brane solutions. One should conclude that there are
no horizons and therefore no black brane solutions. The horizons,
if any exist, are given by the zeros of the function $f(\rho
)=g_{\rho \rho }^{-1}$. Let us denote the zeros of $f(\rho )$ by
$r_{+}$. The function $f(\rho )$ is negative for $\rho <r_{+}$ and
positive for $\rho >r_{+}$, and therefore one may think that the
hypersurface of constant time and $\rho =r_{+}$ is the horizon.
However, this analysis is not correct. Indeed, one may note that
$g_{\rho \rho }$ and $g_{\phi \phi }$ are related by $f(\rho
)=g_{\rho \rho }^{-1}=l^{-2}g_{\phi \phi }$, and therefore when
$g_{\rho \rho }$ becomes negative (which occurs for $\rho <r_{+}$)
so does $g_{\phi
\phi }$. This leads to an apparent change of signature of the metric from $%
(n-1)+$ to $(n-2)+$, and therefore indicates that we are using an incorrect
extension. To get rid of this incorrect extension, we introduce the new
radial coordinate $r$ as
\[
r^{2}=\rho ^{2}-r_{+}^{2}\Rightarrow d\rho ^{2}=\frac{r^{2}}{r^{2}+r_{+}^{2}}%
dr^{2}.
\]
With this new coordinate, the metric (\ref{Metr1a}) is
\begin{eqnarray}
ds^{2} =&&-\frac{r^{2}+r_{+}^{2}}{l^{2}}\left( \Xi dt-ad\phi
\right) ^{2}+f(r)\left( \frac{a}{l}dt-\Xi ld\phi \right) ^{2}
\nonumber\\
&& +\frac{r^{2}}{(r^{2}+r_{+}^{2})f(r)}dr^{2}+\frac{r^{2}+r_{+}^{2}}{l^{2}}%
dX^{2},  \label{Metr1b}
\end{eqnarray}
where the coordinates $r$ and $\phi $ assume the values $0\leq r<\infty $
and $0\leq \phi <2\pi $, and $f(r)$ is now given as
\begin{equation}
f(r)=\frac{r^{2}+r_{+}^{2}}{l^{2}}+\frac{8ml^{n-2}}{%
(r^{2}+r_{+}^{2})^{(n-2)/2}}-\frac{8}{(n-1)(n-2)}\frac{q^{2}l^{2n-4}}{%
(r^{2}+r_{+}^{2})^{n-2}}.  \label{Ff2}
\end{equation}
The gauge potential in the new coordinate is
\begin{equation}
A_{\mu }=\frac{2}{(n-2)}\frac{ql^{(n-3)}}{(r^{2}+r_{+}^{2})^{(n-2)/2}}\left(
a\delta _{\mu }^{0}-\Xi l^{2}\delta _{\mu }^{\phi }\right) .  \label{pot2}
\end{equation}
The function $f(r)$ given in Eq. (\ref{Ff2}) is positive in the whole
spacetime and is zero at $r=0$. Also note that the Kretschmann scalar does
not diverge in the range $0\leq r<\infty $. Therefore this spacetime has no
curvature singularities and no horizons. However, it has a conic geometry
and has a conical singularity at $r=0$, since:
\[
\lim_{r\rightarrow 0}\frac{1}{r}\sqrt{\frac{g_{\phi \phi }}{g_{rr}}}=\frac{nh%
}{2l}+\frac{4q^{2}}{n-1}\left( \frac{l}{h}\right) ^{2n-3}\neq 1.
\]
That is, as the radius $r$ tends to zero, the limit of the ratio ``\textrm{%
circumference/radius}'' is not $2\pi $ and therefore the spacetime
has a conical singularity at $r=0$. Of course, one may ask for the
completeness of the spacetime with $r\geq 0$ (or $\rho \geq
r_{+}$). It is easy to see that the spacetime described by Eq.
(\ref{Metr1b}) is both null and timelike geodesically complete
(see the Appendix for more detail).

\subsection{Longitudinal magnetic field solutions with more rotation
parameters\label{Lmag2}}

The rotation group in $n+1$ dimensions is $SO(n)$ and therefore
the number of independent rotation parameters is $[(n+1)/2]$,
where $[x]$ is the integer part of $x$. We now generalize the
above solution given in Eq. (\ref {Metr1a}) with $k\leq \lbrack
(n+1)/2]$ rotation parameters. This generalized solution can be
written as
\begin{eqnarray}
ds^{2} =&&-\frac{r^{2}+r_{+}^{2}}{l^{2}}\left( \Xi dt-{{\sum_{i=1}^{k}}}%
a_{i}d\phi^{i}\right) ^{2}+f(r)\left( \sqrt{\Xi ^{2}-1}dt-\frac{\Xi }{\sqrt{%
\Xi ^{2}-1}}{{\sum_{i=1}^{k}}}a_{i}d\phi ^{i}\right) ^{2}  \nonumber \\
&&+\frac{r^{2}dr^{2}}{(r^{2}+r_{+}^{2})f(r)}+\frac{r^{2}+r_{+}^{2}}{%
l^{2}(\Xi ^{2}-1)}{\sum_{i<j}^{k}}(a_{i}d\phi _{j}-a_{j}d\phi _{i})^{2}+%
\frac{r^{2}+r_{+}^{2}}{l^{2}}dX^{2},  \label{Metr2}
\end{eqnarray}
where $\Xi=\sqrt{1+\sum_{i}^{k}a_{i}^{2}/l^{2}}$, $dX^{2}$ is the
Euclidean
metric on the $( n-k-1)$-dimensional submanifold and $f(r)$ is the same as $%
f(r)$ given in Eq. (\ref{Ff2}). The gauge potential is
\begin{equation}
A_{\mu
}=\frac{2}{(n-2)}\frac{ql^{(n-2)}}{(r^{2}+r_{+}^{2})^{(n-2)/2}}\left(
\sqrt{\Xi ^{2}-1}\delta _{\mu }^{0}-\frac{\Xi}{\sqrt{\Xi
^{2}-1}}a_{i}\delta _{\mu }^{i}\right);\hspace{.5cm} {\text{(no
sum on i)}}.  \label{Pot2}
\end{equation}
Again this spacetime has no horizon and curvature singularity.
However, it has a conical singularity at $r=0$. One should note
that these solutions are different from those discussed in
\cite{Awad}, which were electrically charged rotating black brane
solutions.

\subsection{Angular magnetic field solutions with one boost parameter\label%
{Amag1}}

The generalization of four-dimensional spacetime with angular
magnetic field to the case of an $n+1$-dimensional AAdS charged
rotating brane with one boost parameter is
\begin{eqnarray}
ds^{2} =&&-\frac{r^{2}+r_{+}^{2}}{l^{2}}\left( \Xi dt-\frac{v}{l}dz\right)
^{2}+f(r)\left( \frac{v}{l}dt-\Xi dz\right) ^{2}  \nonumber \\
&&+\frac{r^{2}dr^{2}}{(r^{2}+r_{+}^{2})f(r)}+(r^{2}+r_{+}^{2})d\Omega ^{2},
\label{Metr3}
\end{eqnarray}
where $\Xi =\sqrt{1+v^{2}/l^{2}}$, $d\Omega
^{2}=\sum_{i=1}^{n-2}(d\phi ^{i})^{2}$, and $f(r)$ is the same as
before given in Eq. (\ref{Ff2}). The
coordinates $r$ and $\phi ^{i}$'s assume the values $0\leq r<\infty $ and $%
0\leq \phi ^{i}\leq 2\pi $, and the gauge potential is given by

\begin{equation}
A_{\mu }=\frac{2}{(n-2)}\frac{ql^{n-3}}{(r^{2}+r_{+}^{2})^{(n-2)/2}}\left(
a\delta _{\mu }^{0}-\Xi l\delta _{\mu }^{z}\right) .  \label{Pot3}
\end{equation}
Using the same arguments given for the case of longitudinal
magnetic field solutions discussed in Sec. II B, one can show that
this spacetime has no curvature singularity, no horizons and no
conical singularity.

\subsection{Angular magnetic field solutions with more boost parameters\label%
{Amag2}}

In this section we introduce the solutions of the Einstein-Maxwell
equation with no rotation parameter and $\kappa$ boost parameters.
The maximum number of boost parameters can be $n-2$. In this case
the solution can be written as
\begin{eqnarray}
ds^{2} =&&-\frac{r^{2}+r_{+}^{2}}{l^{2}}\left( \Xi dt-l^{-1}{{%
\sum_{i=1}^{\kappa}}}v_{i}dx^{i}\right) ^{2}+f(r)\left( \sqrt{\Xi ^{2}-1}dt-%
\frac{\Xi }{l\sqrt{\Xi ^{2}-1}}{{\sum_{i=1}^{\kappa}}}v_{i}dx^{i}\right) ^{2}
\nonumber \\
&&+\frac{r^{2}+r_{+}^{2}}{l^{4}(\Xi ^{2}-1)}\text{ }{\sum_{i<j}^{\kappa}}%
(v_{i}dx_{j}-v_{j}dx_{i})^{2}+\frac{r^{2}dr^{2}}{(r^{2}+r_{+}^{2})f(r)}%
+(r^{2}+r_{+}^{2})d\Omega^{2},  \label{Metr4}
\end{eqnarray}
where $\Xi =\sqrt{1+\sum_{i}^{\kappa}v_{i}^{2}/l^{2}}$, $d\Omega
^{2}=\sum_{i=1}^{n-\kappa-1}(d\phi ^{i})^{2}$, and $f(r)$ is given in Eq. (%
\ref{Ff2}). The gauge potential is given by

\begin{equation}
A_{\mu }=\frac{2}{(n-2)}\frac{\lambda l^{(n-2)}}{(r^{2}+r_{+}^{2})^{(n-2)/2}}%
\left( \sqrt{\Xi ^{2}-1}\delta _{\mu }^{0}-\frac{\Xi}{l\sqrt{\Xi ^{2}-1}}%
v_{i}\delta _{\mu }^{i}\right);\hspace{.5cm}\text{(no sum on i)}.
\label{Pot4}
\end{equation}
Again this spacetime has no curvature singularity, no horizons,
and no conical singularity.

\section{The Conserved Quantities of Magnetic Rotating Brane\label{Cons}}

It is well known that the gravitational action given in Eq.
(\ref{Actg}) diverges. A systematic method of dealing with this
divergence is through the use of the counterterm method inspired
by AdS/CFT correspondence. The AdS/CFT
correspondence states that, if the metric near the conformal boundary ($%
x\rightarrow 0$) can be expanded in the AAdS form,
\begin{equation}
ds^{2}=\frac{dx^{2}}{l^{2}x^{2}}+\frac{1}{x^{2}}\gamma _{ij}dx^{i}dx^{j},
\label{metc}
\end{equation}
with nondegenerate metric $\gamma $, then one may remove the
divergent terms in the action by adding a counterterm action,
$I_{\text{ct}}$, which is a functional of the boundary curvature
invariants. The counterterm for asymptotically AdS spacetimes up
to seven dimensions is
\begin{eqnarray}
I_{\text{ct}} &=&\frac{1}{8\pi }\int_{\partial \mathcal{M}_{\infty }}d^{n}x\sqrt{%
-\gamma }\{\frac{n-1}{l}-\frac{l\Theta (n-3)}{2(n-2)}R  \nonumber \\
&&-\frac{l^{3}\Theta (n-5)}{2(n-4)(n-2)^{2}}\left( R_{ab}R^{ab}-\frac{n}{%
4(n-1)}R^{2}\right) +...\},  \label{Actct1}
\end{eqnarray}
where $R$, $R_{abcd}$, and $R_{ab}$ are the Ricci scalar, Riemann,
and Ricci tensors of the boundary metric $\gamma _{ab}$, and
$\Theta (x)$ is the step function, which is equal to $1$ for
$x\geq 0$ and zero otherwise. Although other counterterms (of
higher mass dimension) may be added to $I_{ct}$, they will make no
contribution to the evaluation of the action or Hamiltonian due to
the rate at which they decrease toward infinity, and we shall not
consider them in our analysis here. These counterterms have been
used by many authors for a wide variety of spacetimes, including
Schwarzschild-AdS, topological Schwarzschild-AdS, Kerr-AdS,
Taub-NUT-AdS, Taub-bolt-AdS, and Taub-bolt-Kerr-AdS \cite{EJM}.

Of course, for even $n$ one has logarithmic divergences in the
partition function which can be related to the Weyl anomalies in
dual conformal field theory \cite{Sken1}. These logarithmic
divergences associated with the Weyl anomalies of the dual field
theory for $n=4$ and $n=6$ are \cite{Sken2}
\begin{eqnarray}
I_{\mathrm{\log }} &=&-\frac{\ln \epsilon }{64\pi l^{3}}\int d^{4}x\sqrt{%
-\gamma ^{0}}\left[ (R_{ij}^{0})R^{(0)ij}-\frac{1}{3}(R^{0})^{2}\right] ,
\label{Ano4} \\
I_{\log } &=&\frac{\ln \epsilon }{8^{4}\pi l^{3}}\int d^{6}x\sqrt{-\gamma
^{0}}\{\frac{3}{50}(R^{0})^{3}+R^{(0)ij}R^{(0)kl}R_{ijkl}^{0}-\frac{1}{2}%
R^{0}R^{(0)ij}R_{ij}^{0}  \nonumber \\
&&\hspace{3cm}+\frac{1}{5}R^{(0)ij}D_{i}D_{j}R^{0}-\frac{1}{2}R^{(0)ij}\Box
^{0}R_{ij}^{0}+\frac{1}{20}R^{0}\Box ^{0}R^{0}\}.  \label{Ano6}
\end{eqnarray}
In Eqs. (\ref{Ano4}) and (\ref{Ano6}) $R^{0}$ and $R^{(0)ij}$ are
the Ricci scalar and Ricci tensor of the leading order metric
$\gamma ^{0}$ in the following expansion:
\begin{equation}
\gamma _{ij}=\gamma _{ij}^{0}+x^{2}\gamma _{ij}^{2}+x^{4}\gamma
_{ij}^{4}+...,  \label{exp}
\end{equation}
and $D_{i}$ is the covariant derivative constructed by the leading
order metric $\gamma ^{0}$. Also, one should note that the
inclusion of matter fields in the gravitational action produces an
additional logarithmic divergence in the action for even $n$. This
logarithmic divergence for $n=4$ and $n=6$ are \cite{Sken2}
\begin{eqnarray}
I_{\log }^{\mathrm{em}} &=&\frac{\ln \epsilon }{64\pi l}\int d^{4}x\sqrt{%
\gamma ^{0}}F^{(0)ij}F_{ij}^{0},  \label{logem4} \\
I_{\log }^{\mathrm{em}} &=&\frac{\ln \epsilon }{8\pi l^{3}}\int d^{6}x\sqrt{%
-\gamma ^{0}}\{\frac{1}{16}R^{0}F^{(0)ij}F_{ij}^{0}-\frac{1}{8}%
R^{(0)ij}F_{i}^{(0)l}F_{jl}^{0}  \nonumber \\
&&\hspace{3.5cm}+\frac{1}{64}%
F^{(0)ij}(D_{j}D^{k}F_{ki}^{0}-D_{i}D^{k}F_{kj}^{0})\}.  \label{logem6}
\end{eqnarray}
where $F_{ij}^{0}$ is the leading term of the electromagnetic
field on the conformal boundary. All the contractions in Eqs.
(\ref{logem4}) and \ref
{logem6}) should be done by the leading order metric $\gamma _{ij}^{0}$. In $%
4<n<7$ the matter field will cause a power law divergence in the
action, which can be removed by a counterterm of the form
\cite{Sken2}
\begin{equation}
I_{ct}^{\mathrm{em}}=\frac{1}{256\pi }\int d^{n}x\sqrt{_{-}\gamma }\frac{%
(n-8)}{(n-4)}\Theta (n-5)F^{ij}F_{ij}.  \label{Actct2}
\end{equation}
Thus, the total action can be written as a linear combination of the gravity
term (\ref{Actg}) and the counterterms (\ref{Actct1}) and (\ref{Actct2}).
For the charged rotating and travelling magnetic branes investigated in this
paper the counterterm $I_{\mathrm{ct}}^{\mathrm{em}}$ is zero and therefore
the total action is
\begin{equation}
I=I_{G}+I_{\mathrm{ct}}.  \label{Acttot}
\end{equation}
Having the total action, one can use the Brown and York definition
\cite {Brown} to construct a divergence-free stress-energy tensor
as
\begin{eqnarray}
T^{ab} &=&\frac{1}{8\pi }\{(K^{ab}-K\gamma ^{ab})-\frac{n-1}{l}\gamma ^{ab}+%
\frac{l}{n-2}(R^{ab}-\frac{1}{2}R\gamma ^{ab})  \nonumber \\
&&\ +\frac{l^{3}\Theta (n-5)}{(n-4)(n-2)^{2}}[-\frac{1}{2}\gamma
^{ab}(R^{cd}R_{cd}-\frac{n}{4(n-1)}R^{2})-\frac{n}{(2n-2)}RR^{ab}  \nonumber
\\
&&\ +2R_{cd}R^{acbd}-\frac{n-2}{2(n-1)}\nabla ^{a}\nabla ^{b}R+\nabla
^{2}R^{ab}-\frac{1}{2(n-1)}\gamma ^{ab}\nabla ^{2}R]+...\}.  \label{Stres}
\end{eqnarray}
The above stress tensor is divergence-free for $n\leq 6$, but we
can always add more counterterms to have a finite action in higher
dimensions (see, e.g., \cite{Krau}).

The conserved charges associated to a Killing vector $\xi ^{a}$ is
\begin{equation}
\mathcal{Q}(\xi )=\int_{\mathcal{B}}d^{n}x\sqrt{\sigma }n^{a}T_{ab}\xi ^{b},
\label{Con}
\end{equation}
where $\sigma $ is the determinant of the metric $\sigma _{ij}$,
appearing in the Arnowitt-Deser-Misner-like decomposition of the
boundary metric,
\begin{equation}
ds^{2}=-N^{2}dt^{2}+\sigma _{ij}(dx^{i}+N^{i}dt)(dx^{j}+N^{j}dt).
\label{ADM}
\end{equation}
In Eq. (\ref{ADM}), $N$ and $N^{i}$ are the lapse and shift
functions, respectively. For boundaries with timelike Killing
vector ($\xi =\partial /\partial t$), rotational Killing vector
field $(\zeta _{i}=\partial /\partial \phi ^{i})$, and
translational Killing vector $(\varsigma _{i}=\partial /\partial
x^{i})$ one obtains the conserved mass, angular
momentum and linear momentum of the system enclosed by the boundary $%
\mathcal{B}$. In the context of AdS/CFT correspondence, the limit
in which the boundary $\mathcal{B}$ becomes infinite
$(\mathcal{B}_{\infty })$ is taken, and the counterterm
prescription ensures that the action and conserved charges are
finite. No embedding of the surface $\mathcal{B}$ into a reference
spacetime is required and the quantities which are computed are
intrinsic to the spacetimes.

For our case, horizonless rotating spacetimes, the first Killing vector is $%
\xi =\partial /\partial t$ and therefore its associated conserved
charge is the total mass of the system enclosed by the boundary
given by
\begin{equation}
M=\int_{\mathcal{B}}d^{n-1}x \sqrt{\sigma }T_{ab}n^{a}\xi ^{b}=\frac{V_{n-1}%
}{8\pi }ml^{p-1}\left[ n(\Xi ^{2}-1)+1\right] ,  \label{Mas}
\end{equation}
where $V_{n-1}$ denotes the volume of the hypersurface boundary
$\mathcal{B}$ at constant $t$ and $r$, and $p$ is the number of
angular coordinates $\phi ^{i}$ of the spacetime. One may note
that in the case of a spacetime with a longitudinal magnetic
field, the numbers of angular coordinates and rotation parameters
are the same, but for a spacetime with an angular magnetic field,
they are different.

For the case of spacetimes with a longitudinal magnetic field, the
charges associated with the rotational Killing symmetries
generated by $\zeta _{i}=\partial /\partial \phi ^{i}$ are the
components of total angular momentum of the system enclosed by the
boundary given as
\begin{equation}
J_{i}=\int_{\mathcal{B}}d^{n-1}x \sqrt{\sigma }T_{ab}n^{a}\zeta _{i}^{b}=%
\frac{V_{n-1}}{8\pi }n\Xi l^{p-1}ma_{i}.  \label{Ang}
\end{equation}

In the case of the spacetimes with an angular magnetic field
introduced in Sec. II E, we encounter conserved quantities
associated with translational Killing symmetries generated by
$\varsigma _{i}=\partial /\partial x^{i}$. These conserved
quantities are the components of linear momentum calculated as

\begin{equation}
P_{i}=\int_{\mathcal{B}}d^{n-1}x \sqrt{\sigma }T_{ab}n^{a}\varsigma _{i}^{b}=%
\frac{V_{n-1}}{8\pi }n\Xi l^{p-2}mv_{i}.  \label{Lin}
\end{equation}

Now we show that the $r$ divergence of the total action can be
removed by the counterterm method. We first calculate the
logarithmic divergences due
to the Weyl anomaly and matter field given in Eqs. (\ref{Ano4}), (\ref{Ano6}%
), (\ref{logem4}), and (\ref{logem6}). The leading metric
$h_{ij}^{0}$ can be obtained as
\begin{equation}
h_{ij}^{0}dx^{i}dx^{j}=-\frac{1}{l^{2}}dt^{2}+d\Omega ^{2}+dX^{2}.
\label{gamma0}
\end{equation}
Therefore the curvature scalar $R^{0}(h^{0})$ and Ricci tensor $%
R_{ij}^{0}(h^{0})$ are zero. Also, it is easy to show that
$F_{ij}^{0}$ in Eqs. (\ref{logem4}) and (\ref{logem6}) vanishes.
Thus, all the logarithmic divergences for the $(n+1)$-dimensional
solutions given in Sec. \ref{Sol} are zero. It is also a matter of
calculation to show that the counterterm action due to the
electromagnetic field in Eq. (\ref{Actct2}) is zero. Using
Eqs. (\ref{Actg}), (\ref{Actct1}), and (\ref{Acttot}), one can show that the $%
r$ divergence of the action will be removed.

Next, we calculate the electric charge of the solutions. To
determine the electric field we should consider the projections of
the electromagnetic field tensor on special hypersurfaces. The
normal to such hypersurfaces for the spacetimes with a
longitudinal magnetic field is
\[
u^{0}=\frac{1}{N},\text{ \ }u^{r}=0,\text{ \ }u^{i}=-\frac{N^{i}}{N},
\]
and the electric field is $E^{\mu }=g^{\mu \rho }F_{\rho \nu
}u^{\nu }$. Then the electric charge $Q$ can be found by
calculating the flux of the electromagnetic field at infinity,
yielding
\begin{equation}
Q=\frac{V_{n-1}}{4\pi }\sqrt{\Xi ^{2}-1}l^{p-2}q.  \label{elecch}
\end{equation}
Note that the electric charge is proportional to the rotation
parameter or boost parameter, and is zero for the case of a static
solution.

\section{CLOSING REMARKS}

In this paper, we introduced two classes of solutions of
Einstein-Maxwell gravity which are asymptotically anti-de Sitter.
The first class of solutions yields a rotating spacetime with a
longitudinal magnetic field. We found that these solutions have no
curvature singularity and no horizons, but have conic singularity
at $r=0$. In these spacetimes, when all the rotation parameters
are zero (static case), the electric field vanishes, and therefore
the brane has no net electric charge. For the spinning brane, when
one or more rotation parameters are nonzero, the brane has a net
electric charge which is proportional to the magnitude of rotation
parameter given by $\sum_{i}^{k}a_{i}^{2}$. The second class of
solutions yields a spacetime with angular magnetic field. These
solutions have no curvature singularity, no horizon, and no conic
singularity. Again we found that the branes in these spacetimes
have no net electric charge when all the boost parameters are
zero. We also showed that, for the case of traveling branes with
one or more nonzero boost parameters, the net electric charge of
the
brane is proportional to the magnitude of the velocity of the brane ($%
\sum_{i}^{\kappa}v_{i}^{2}$).

We also used the counterterm method inspired by the AdS/CFT
correspondence conjecture and calculated the conserved quantities
of the two classes of solutions. We found that the logarithmic
divergencies associated with the Weyl anomalies and matter field
are zero, and showed that the $r$ divergence of the action is
removed by use of the counterterm method.
\bigskip
\begin{center}
{\bf APPENDIX}
\end{center}

In this appendix, we want to show that the spacetime described by
the metric (\ref{Metr1b}) is geodesically complete for $r\geq 0$
\cite{Lem2,Hor}. In fact, we want to show that every null or
timelike geodesic starting from an arbitrary point can either
extend to infinite values of the affine parameter along the
geodesic or end on a singularity at $r=0$. To do this, we first
perform the rotation boost $(\Xi t-a \phi) \mapsto t;
\hspace{.1cm} (a t-\Xi l^2 d\phi) \mapsto l^2 d\phi$ in the
$t-\phi$ plane. Then the metric (\ref{Metr1b}) becomes
\[
ds^{2}=-\frac{r^{2}+r_{+}^{2}}{l^{2}}dt^{2}+\frac{r^{2}}{%
(r^{2}+r_{+}^{2})f(r)}dr^{2}+l^{2}f(r)d\phi ^{2}+\frac{r^{2}+r_{+}^{2}}{l^{2}%
}dX^{2}.
\]
Using the geodesic equation, one obtains
\begin{eqnarray}
&& \dot{t}=\frac{l^2}{r^2+r_{+}^2}E, \hspace{.5cm}
\dot{x^i}=\frac{l^2}{r^2+r_{+}^2}P^i,
\hspace{.5cm}\dot{\phi}=\frac{1}{l^2f(r)}L, \nonumber\\
&& r^{2} \dot{r}^{2}=(r^{2}+r_{+}^{2})f(r)\left[ \frac{l^{2}(E^{2}-\textbf{P}^{2})}{%
r^{2}+r_{+}^{2}}-\alpha
\right]-\frac{r^{2}+r_{+}^{2}}{l^{2}}L^{2}, \nonumber
\end{eqnarray}
where the overdot denotes the derivative with respect to an affine
parameter and $\alpha$ is zero for null geodesics and  $+1$ for
timelike geodesics. $E$, $L$, and $P^{i}$ are the conserved
quantities associated with the coordinates $t$, $\phi$, and
$x^{i}$ respectively, and
$\textbf{P}^2=\sum_{i=1}^{n-2}(P^{i})^{2}$. Notice that $f(r)$ is
always positive for $r>0$ and zero for $r=0$.

First we consider the null geodesics ($\alpha=0$). $(\texttt{i})$
If $E^{2}>\textbf{P}^{2}$ the
spiraling particles ($L>0$) coming from infinity have a turning point at $%
r_{tp}>0$, while the nonspiraling particles ($L=0$) have a turning point at $%
r_{tp}=0$. $(\texttt{ii})$ If $E=\textbf{P}$ and $L=0$, whatever
is the value of $r$, $\dot{r}$ and  $\dot{\phi}$, vanish and
therefore the null particles move in a straight line in the
$(n-2)$-dimensional submanifold spanned by $x^{1}$ to $x^{n-2}$.
$(\texttt{iii})$ For $E=\textbf{P}$ and $L\neq0$, and also for
$E^{2}<\textbf{P}^{2}$ and any value of $L$, there is no possible
null geodesic.

Now, we analyze the timelike geodesics ($\alpha=+1$). A timelike
geodesic is
possible only if $l^{2}(E^{2}-\textbf{P}^{2})>r_{+}^{2}$. In this case spiraling ($%
L\neq0$) timelike particles are bound between $r_{tp}^{a}$ and
$r_{tp}^{b}$ given by
\[
0<r_{tp}^{a}\leq
r_{tp}^{b}<\sqrt{l^{2}(E^{2}-\textbf{P}^{2})-r_{+}^{2}},
\]
while the turning points for the nonspiraling particles ($L=0$)
are $r_{tp}^{1}=0$ and
$r_{tp}^{2}=\sqrt{l^{2}(E^{2}-\textbf{P}^{2})-r_{+}^{2}}$. Thus,
we have confirmed that the spacetime described by Eq.
(\ref{Metr1b}) is both null and timelike geodesically complete.

\end{document}